\documentclass[apjl]{emulateapj}
\submitted{C. Winteler et al. 2012 ApJ 750 L22}
\usepackage{graphicx}
\usepackage{natbib}
\usepackage{amsmath}

\begin{document}

% title
\title{Magnetorotationally driven Supernovae as the origin of \\ early galaxy $r$-process elements?}
\shorttitle{$r$-process in 3D MHD Jets}

% authors
\author{C. Winteler \altaffilmark{1},
  R. K\"appeli \altaffilmark{2},
  A. Perego \altaffilmark{1},
  A. Arcones \altaffilmark{3,4},
  N. Vasset \altaffilmark{1},
  N. Nishimura \altaffilmark{1},\\
  M. Liebend\"orfer \altaffilmark{1} and
  F.-K. Thielemann \altaffilmark{1}
  }

\shortauthors{Winteler et al.}

% affiliations
\altaffiltext{1}{Physics Department,
University of Basel,
Klingelbergstrasse 82,
CH-4056 Basel,
Switzerland}

\altaffiltext{2}{Seminar for applied Mathematics,
ETH Z\"urich,
R\"amistrasse 101,
8092 Z\"urich,
Switzerland}

\altaffiltext{3}{Institut f\"ur Kernphysik, 
Technische Universit\"at Darmstadt, 
Schlossgartenstra{\ss}e 2, 
D-64289 Darmstadt, 
Germany}

\altaffiltext{4}{GSI  Helmholtzzentrum 
f\"ur Schwerionenforschung GmbH, 
Planckstr. 1, 
D-64291 Darmstadt, 
Germany}

% corresponding authors
\email{christian.winteler@unibas.ch}

\begin{abstract}
We examine magnetorotationally driven supernovae as sources of $r$-process elements in the early Galaxy. On the basis of thermodynamic histories of tracer particles from a three-dimensional magnetohydrodynamical core-collapse supernova model with approximated neutrino transport, we perform nucleosynthesis calculations with and without considering the effects of neutrino absorption reactions on the electron fraction ($Y_{e}$) during post-processing. We find that the peak distribution of $Y_{e}$ in the ejecta is shifted from $\sim0.15$ to $\sim0.17$ and broadened toward higher $Y_{e}$ due to neutrino absorption. Nevertheless, in both cases the second and third peaks of the solar $r$-process element distribution can be well reproduced. The rare progenitor configuration that was used here, characterized by a high rotation rate and a large magnetic field necessary for the formation of bipolar jets, could naturally provide a site for the strong $r$-process in agreement with observations of the early galactic chemical evolution. 
\end{abstract}


%
%________________________________________________________________

\section{Introduction}
\label{sec:intro}
The supernova (SN) mechanism explaining how massive stars end in a central collapse to a neutron star (or a black hole) and explosive ejection of the outer layers is still debated. It has been related to neutrino emission from the hot collapsed core and accreted matter \citep{1990RvMP...62..801B}, and this (regular) scenario is again becoming more and more promising with recent multi-D hydrodynamic approaches and improved (spectral) neutrino transport \citep[e.g., ][]{2010ApJS..189..104M,2012ApJ...747...73L, 2010PThPS.186...87L,2011CoPhC.182.1764B}. The equation of state (EoS) caused explosion scenario has found some recent revival, based on the quark-hadron phase transition at supernuclear densities \citep{2009PhRvL.102h1101S}, but its working depends on specific choices of EoS properties in a very narrow parameter range. Other options are related to rotation and magnetic fields, a topic which has been discussed for more than 30 years, but required three-dimensional (3D) modeling and had therefore only been addressed with limited success in the early days. The major outcome was that high rotation rates and (possibly unrealistically high) magnetic fields were required to launch explosions. The question is whether such magnetic fields can be attained during collapse with rotation and on which timescale after collapse \citep[and references therein]{1970ApJ...161..541L,1976ApJ...204..869M, 1976Ap&SS..41..321B,1979A&A....80..147M}.  This topic has recently been re-addressed by \citet{2008IJMPD..17.1411M} and \citet{2009ApJ...691.1360T}. In the present Letter we want to present results obtained with our 3D magneto-hydrodynamics code FISH \citep{2011ApJS..195...20K}.

There is another reason why such magnetohydrodynamically driven explosions are of possible interest: the search for the site of a (full) $r$-process early in galactic evolution. The initially postulated neutrino wind in regular (neutrino-driven) supernovae \citep{1997ApJ...482..951H,1997ApJS..112..199M,2010ApJ...712.1359F} (a) did not result in having sufficiently high entropies \citep{2001ApJ...562..887T,2006ApJ...650L..79W,2007A&A...467.1227A}, and (b) in addition, the innermost ejecta turned out to be proton-rich instead of neutron-rich \citep{2006ApJ...644.1028P,2006PhRvL..96n2502F}, a situation which has now also been shown in long-term simulations \citep{2010A&A...517A..80F,2010PhRvL.104y1101H}. Electron-capture supernovae, which explode without a long phase of accretion onto the proto-neutron star (PNS), apparently provide more favorable conditions \citep{2011ApJ...726L..15W}. However, also the proton to nucleon ratios obtained under such conditions do not support a strong $r$-process, which successfully reproduces the platinum peak of r-elements around  $A = 195$. A similar result was recently obtained for core-collapse supernova (CCSN) explosions triggered  by a quark-hadron phase transition during the early post-bounce phase when investigating their detailed nucleosynthesis \citep{2011arXiv1112.5684N}. Both types of events might support a weak but not a full $r$-process. 

Neutron star mergers have been shown to be powerful sources of $r$-process matter, in fact ejecting a  factor of 100 to 1000 more $r$-process material than required on average from CCSNe,  if those would have to explain solar $r$-process element abundances. This would actually support the large scatter  of Eu/Fe found in very metal-poor stars. The only problem is that it might be hard to explain the early appearance of $r$-process matter for metallicities at and below [Fe/H] = $--3$ \citep{2004A&A...416..997A}. Some recent studies, which include the fact that our Galaxy is  possibly the result from smaller merging subsystems (with different star formation rates) have been  expected to show a way out of this dilemma. If this cannot be solved, we need another strong $r$-process source already at low metallicities, and possibly jets from rotating core collapses with strong magnetic fields could be the solution \citep{2003ApJ...587..327C,2006ApJ...642..410N,2008ApJ...680.1350F}. 

The present Letter has the aim to explore the results from our 3D magnetohydrodynamic (MHD) calculations, which lead to bipolar jet ejection. The following Section \ref{sec:3D MHD-CCSN} will discuss the initial models and the explosion dynamics; Section \ref{sec:nucleosynth} will present nucleosynthesis results. Section \ref{sec:discussion_outlook} is devoted to a discussion of uncertainties and an outlook on future investigations.

%__________________________________________________________________

\section{3D MHD-CCSN model}
\label{sec:3D MHD-CCSN}
The calculation presented here was performed with the computational setup similar to our previous investigations \citep{2005NuPhA.758...59L,2010A&A...514A..51S}. The initially innermost (600 km)$^3$ of the massive star are covered by a 3D Cartesian domain uniformly discretized by $600^3$ cells, resulting in a 1 km resolution, that is embedded in a spherically symmetric domain encompassing the iron core and parts of the silicon shell.  The magnetic fluid is evolved with the \texttt{FISH} code \citep{2011ApJS..195...20K}, solving the ideal MHD equations. The spherically symmetric domain is evolved with the \texttt{AGILE} code \citep{2002ApJS..141..229L}. The gravitational potential is approximated by an effective axisymmetric mass distribution that includes general relativistic monopole corrections \citep{2006A&A...445..273M}. We use the \citet{1991NuPhA.535..331L} EoS with nuclear compressibility 180 MeV. We have included a Lagrangian component in the form of tracer particles which are passively advected with the flow. They record the thermodynamic conditions of a particular fluid element and serve as input to the post-processing nucleosynthesis calculations.

The transport of the electron neutrinos and anti-neutrinos is approximated  by a 3D spectral leakage scheme, based on previous gray leakage schemes \citep[and references therein]{2003MNRAS.342..673R}. The neutrino energy is discretized with 12 geometrically increasing energy groups spanning the range $E_\nu = 3 -- 200$ MeV. The amount of energy and particles locally released is calculated for each bin as an interpolation between the diffusive rates and the (free streaming) production rates, depending on the local neutrino optical depth. For the computation of the spectral optical depth we have used a ray-by-ray axisymmetric approximation, calculated on a polar grid encompassing the  full 3D Cartesian domain discretized uniformly with 1km radial spacing and 30 angular rays covering the full $[0,\pi]$ realm. All fundamental neutrino reactions have been included (neutrino scattering on nucleons and nuclei, neutrino absorption/emission on nucleons and nuclei), providing detailed spectral emissivities and opacities \citep{1985ApJS...58..771B}. Inside the neutrinosphere, weak equilibrium is assumed and trapped neutrinos are modeled accordingly; outside of it, no explicit absorption is considered. Thus we can only follow neutrino emission and the associated neutronization of matter. However, the up to now microphysically most complete two-dimensional axisymmetric study of MHD-CCSN with multi-group flux-limited diffusion neutrino transport performed by \citet{2007ApJ...664..416B} has shown, that neutrino heating contributes only $10\%--25\%$ to the explosion energy and is therefore subdominant. This justifies our pragmatic approach at first.

We employed the pre-collapse $15 M_\odot$ model of \citet{2005ApJ...626..350H}. Although the model provides profiles for rotation and magnetic fields, we use an analytic prescription for their distributions and we will comment on this choice in Section \ref{sec:discussion_outlook}. The initial rotation law was assumed to be shellular with $\Omega(r) = \Omega_{0} R_0^2/(r^2+R_0^2)$, $\Omega_0 = \pi$ s$^{-1}$ and $R_0 = 1000$ km corresponding to an initial ratio of rotational energy to gravitational binding energy $T_{\mathrm{rot}}/|W| = 7.63 \times 10^{-3}$. For the magnetic field we have assumed a homogeneous distribution of a purely poloidal field throughout the computational domain of strength $5 \times 10^{12}$ G corresponding to an initial ratio of magnetic energy to gravitational binding energy $T_{\mathrm{mag}}/|W| = 2.63 \times 10^{-8}$.

\begin{figure}[t!]
\epsscale{1.0}
\plotone{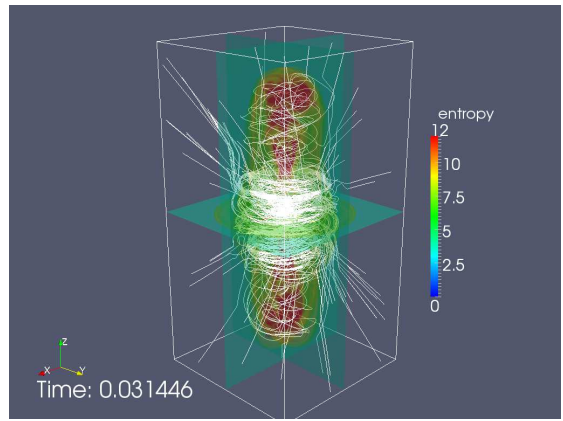}
  \caption{3D entropy contours spanning the coordinates planes with magnetic field lines (white lines) of the MHD-CCSN simulation $\sim 31$ ms after bounce. The 3D domain size is $700 \times 700 \times 1400$ km.
  }
  \label{fig:fig01}
\end{figure}

The computed model then undergoes gravitational collapse and experiences core-bounce due to the stiffening of the EoS above nuclear saturation density. Conservation of angular momentum in combination with the collapse leads to a massive spin-up of the core, reaching $T_{\mathrm{rot}}/|W| = 6.81 \times 10^{-2}$ at bounce, and significant rotationally induced deformations. During the collapse the magnetic field is amplified by magnetic flux conservation reaching a central strength of $\sim 5 \times 10^{15}$ G and $T_{\mathrm{mag}}/|W| = 3.02 \times 10^{-4}$ at bounce. After bounce, differential rotation winds up the poloidal field very quickly into a very strong toroidal field increasing the magnetic energy/pressure at the expense of rotational energy. Consequently, strongly magnetized regions appear near the rotational axis with an associated magnetic pressure quickly reaching and exceeding that of the local gas pressure. The Lorentz force then becomes dynamically important and matter near the rotational axis is lifted from the PNS and drives a bipolar outflow, i.e., jets are launched. The jets rapidly propagate along the rotational axis and quickly reach the boundary of the initial 3D domain. In order to follow the jet propagation further, we have continuously extended the 3D domain to a final size of $700 \times 700 \times 1400$ km at $\sim 31$ ms after bounce. Figure~\ref{fig:fig01} displays a snapshot at the final time.

The quickly expanding bipolar jets transport energy and neutron rich material outward against the gravitational attraction of the PNS. We have estimated the ejected mass $M_\textrm{ej} = 6.72 \times 10^{-3} M_\odot$ and explosion energy $E_\textrm{exp} = 8.45 \times 10^{49}$ erg by summing over the fluid cells that are gravitationally unbound. We defined a fluid cell as unbound if its total specific energy (internal+kinetic+magnetic+potential) is positive and if the radial velocity is pointing outward. These are admittedly crude lower bound estimates and these numbers were still growing at the end of the simulation.

\section{Nucleosynthesis}
\label{sec:nucleosynth}
The nucleosynthesis calculations are performed with a new extended reaction network \citep{Winteler2011} which represents an advanced (numerically and physically) update of the BasNet network (see, e.g., \citet{2011PrPNP..66..346T}). We use the reaction rates of \citet[][for the FRDM mass model]{2000ADNDT..75....1R}. We use the same weak interaction rates (electron/positron captures and $\beta$-decays) as in \citet{2011PhRvC..83d5809A}. Additionally, we include neutron capture and neutron induced fission rates following \citet{2010A&A...513A..61P} and $\beta$-delayed fission probabilities as described in \citet{2005NuPhA.747..633P}.

\begin{figure}[t!]
\epsscale{1.0}
\plotone{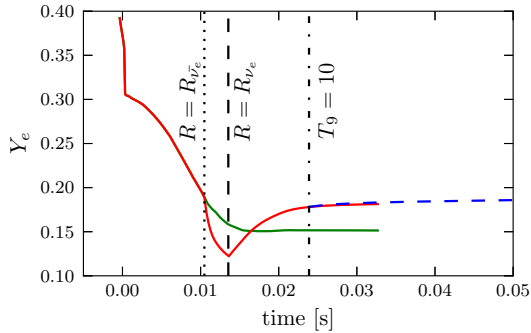}
\caption{Time evolution of $Y_e$ from a single tracer in green (red) for the original data (with neutrino absorption estimate). The blue dashed line represents $Y_e$ evolved by the network. Vertical dotted (dashed) lines correspond to the time when $\bar{\nu}_e$ ($\nu_e$) neutrinosphere is crossed; the dot-dashed vertical line when $T=10\,{\rm GK}$.}
\label{fig:fig02}
\end{figure}

The tracer particles obtained from the simulation provide density, temperature, and electron fraction for the nuclear network, as well as position and velocity, from the beginning to the end of the simulation ($t=t_{f}$). 
After $t_f$, thermodynamic variables are evolved following the prescription in \citet{2008ApJ...680.1350F}. \citet{2006ApJ...642..410N} have shown that the details of the expansion only have a minor impact on the final abundances. For the post-processing we only consider gravitationally unbound tracer particles (see Section \ref{sec:3D MHD-CCSN}). In order to obtain mass integrated abundances we distribute the total ejected mass equally among all ejected tracers. It could be shown that this yields very similar results to post-processing calculations based on the conditions in the unbound cells at the final time \citep{Winteler2011}.

The electron fraction is a key input for the nucleosynthesis and strongly depends on details of the challenging neutrino transport. Although neutrino absorption is crucial to determine the $Y_e$, it is not yet included in the hydrodynamical simulations (where it is expected to have a minor impact on the dynamics; see Section \ref{sec:3D MHD-CCSN}). Therefore, we present two different nucleosynthesis calculations: (1) $Y_e$ is taken from the original tracer particles, (2) the effects of neutrino absorption on $Y_e$ are included also in the network. In this second approach, we post-process the data from the tracer and use the neutrino information obtained with the leakage scheme (Section \ref{sec:3D MHD-CCSN}). We use integrated neutrino luminosities to update the electron fraction outside the neutrinosphere (neglecting the effect of neutrino energy deposition on the matter temperature). The electron fraction of the tracer is evolved using approximated rates for the neutrino emission and absorption on nucleons (see, e.g., \citet{2001A&A...368..527J}).
 
In both of our approaches for $Y_e$, the network calculations start when the temperature decreases to $10$~GK. In the case with post-processing corrections for neutrino absorption, we also consider neutrino reactions from \citet{2006PhRvL..96n2502F} in the nucleosynthesis network. The required neutrino luminosities and mean energies are given by the leakage scheme and assumed to be constant and equal to their values at $t=t_f$.

The evolution of $Y_e$ is presented in Figure~\ref{fig:fig02} for the original simulation data and for the estimate of neutrino absorption. In the latter, high energy $\bar{\nu}_e$ captures on protons decrease $Y_e$ outside the $\bar{\nu}_e$-neutrinosphere. However, beyond the $\nu_e$-neutrinosphere, $\nu_e$ absorption on neutrons dominates and $Y_e$ increases. The fast expansion and the relatively low neutrino mean energy limit the effect of the absorption. This trend is confirmed also by the network for $T \lesssim 10$~GK. 

\begin{figure}[t!]
\epsscale{1.0}
\plotone{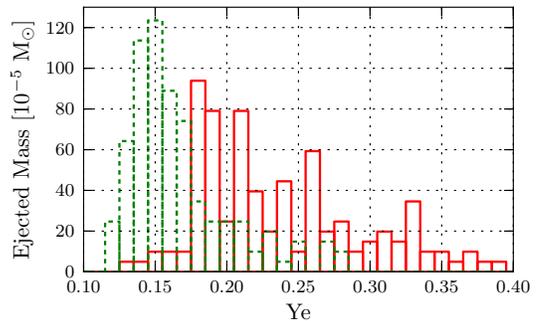}
\caption{Ejecta masses vs. $Y_{e}$ for the original simulation without neutrino captures (green) and including a simplified prescription for neutrino heating (red). The width of a $Y_{e}$ bin is chosen to be $\Delta Y_{e} = 0.01$.}
\label{fig:fig03}
\end{figure}

\begin{figure}[t!]
\epsscale{1.0}
\plotone{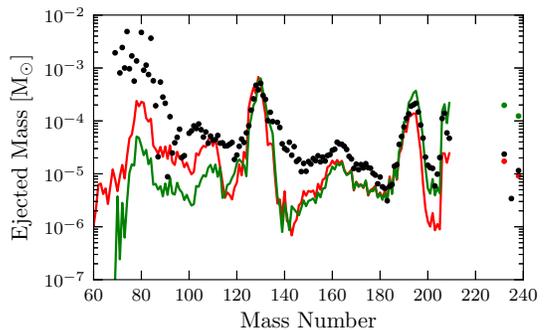}
\caption{Integrated mass fractions for nucleosynthesis calculations with (red)and without (green) neutrino heating. Black dots represent solar $r$-process element abundances \citep{2008ARA&A..46..241S} scaled to fit the red line at $A=130$.}
\label{fig:fig04}
\end{figure}

Figure~\ref{fig:fig03} shows the ejected mass as a function of $Y_e$ for the original simulation data and for the case including neutrino absorption. These corrections shift the peak distribution from $\sim 0.15$ to $\sim 0.17$ and broaden it toward higher $Y_e$. In both approaches, at the onset of the nucleosynthesis, the density is still relatively high, $\rho \approx 10^9 \mathrm{g}\mathrm{cm}^{-3}$, and the electron fraction rather low, $Y_e \approx 0.15 -- 0.3$. This leads to an initial nuclear statistical equilibrium composition rich in neutrons and neutron-rich nuclei. Such conditions are closer to neutron star mergers than to the high entropy wind, which is characterized by an alpha-rich freeze-out with few nuclei close to stability and small amount of neutrons.

Mass integrated abundances are presented in Figure~\ref{fig:fig04} for the two treatments of the electron fraction and compared to solar $r$-process abundances (e.g., \cite{2008ARA&A..46..241S}). Nuclei around the second ($A=130$) and third ($A=195$) $r$-process peaks, as well as the Pb region, can be synthesized in the jets. Note that the improvements in the electron fraction calculation result only in minor variations of the abundances. Moreover, the position of the peaks nicely agrees with solar system abundances. However, the agreement is not equally good for all mass numbers as the nuclei between peaks are slightly underproduced. Note also that the large trough in the mass range $140--160$ is mainly due to the strong $N = 82$ shell-closure of the FRDM mass model. Similar deficiencies were discussed by \citet{2006ApJ...642..410N} and can be cured by using a different mass model. Nuclei in the region below the second peak are produced mainly by charged-particle reactions which occur only in the tracer particles with relatively high electron fraction. This also explains the lower abundances in the range $A=80--100$ for the case of $Y_e$ without neutrino absorption (Figure~\ref{fig:fig04}). This calculation reaches lower $Y_e$ which leads to higher neutron densities and a more abundant third peak and Pb region.

\section{Discussion and Outlook}
\label{sec:discussion_outlook}
Magneto-rotationally driven supernovae suffer one main critique point: the simultaneous presence of fast rotation and strong magnetic fields in the progenitor before collapse. Recent progenitor models computed by \citet{2005ApJ...626..350H} suggest that these special conditions are not reached in common evolutionary paths of massive stars. However, the low-metallicity models of \citet{2006ApJ...637..914W} indicate that for a fraction, perhaps 1\%, of all massive stars, favorable conditions can appear under special circumstances. This rarity of progenitors with these special initial conditions can also be put into accordance with the observed scatter of $r$-process element abundances at low metallicity, combined with the high $r$-process production of $\sim 6 \times 10^{-3}\,M_\odot$ per event. This indicates that these elements have been produced in relatively scarce but efficient events \citep{2008ARA&A..46..241S}. Therefore, the rare progenitor configurations used here could naturally provide a strong $r$-process site in agreement with observations of the early galaxy chemical evolution. 

Currently, the aim to perform long-term global 3D simulations of CCSNe, including possibly a sophisticated radiative transfer of neutrinos, and the aim to simulate the local flow instabilities, leading to magnetic field amplification, seem to be mutually exclusive \citep{2009A&A...498..241O}. Given this difficulty, we followed here the common approach in the literature by taking sufficiently strong initial magnetic fields. By doing so, it is hypothesized that there is a physical process that can sufficiently quickly amplify the magnetic field to dynamic importance by extracting the free rotational energy in differential rotation. This motivates our choice of the high initial poloidal field strength, which by flux compression and rotational winding leads to magnetic fields whose magnitude roughly agree with those expected from the magnetorotational instability at saturation \citep{2009A&A...498..241O}.

There are additional shortcomings in the treatment presented here. This includes that thermodynamic properties of tracer particles are only extrapolated beyond the end of the MHD calculation and the nucleosynthesis results were not yet tested with several nuclear mass models. But the main outcome of this investigation is that full 3D calculations can support the emergence of bipolar jets and that these are not artifacts of up to now axisymmetric approaches \citep{2006ApJ...642..410N,2008ApJ...680.1350F}. Such explosions, resulting from the individual evolution of massive stars rather than complicated binary histories of neutron star mergers, could explain strong $r$-process features during early galactic evolution as observed in many low metallicity stars. The $r$-process production is very efficient, a factor of 100 more than expected on average from supernovae, if they would be responsible for solar $r$-process abundances. The question remains: whether magnetars --- with the magnetic fields required for this outcome --- result from about 1\% of all core collapses or are rarer events. Apparently present knowledge permits this option \citep{1998Natur.393..235K,2009IAUS..259..485K}.

The magnetorotationally driven supernova simulation presented here provides a scenario for a strong $r$-process site seemingly consistent with observations of the early chemical evolution of our Galaxy. Future studies should investigate the effects of the progenitor configuration (like mass, magnetic field and rotation rate), the nuclear input (in particular the mass formula), and a more consistent treatment of neutrino interactions on the robustness of the conclusions drawn here.

% acknowledgements
\acknowledgements
We thank T. Rauscher and I. Panov for useful discussions and help with reaction rates and fission. We acknowledge support and funding by the Swiss National Science Foundation (SNSF). A.A. was supported by Feodor Lynen Fellowship (Humboldt Foundation) and by the Helmholtz-University Young Investigator grant no. VH-NG-825. The authors are additionally supported by EuroGENESIS, a collaborative research program of the European Science Foundation (ESF), CompStar, a research networking program of the ESF, and the FP7 ENSAR project, funded by the European Commission. This work was supported by a grant from the Swiss National Supercomputing Centre (CSCS) under project ID s332.

% bibliography
%\bibliographystyle{references/apj}
%\bibliography{references/apj-jour,references/references}

\end{document}